\documentclass[12pt,a4paper]{article}
\usepackage[utf8]{inputenc}
\usepackage[T1]{fontenc}
\usepackage{amsmath}
\usepackage{amsfonts}
\usepackage{authblk}
\usepackage{amssymb}
\usepackage{graphicx}
\newcommand{\be}{\begin{equation}}
\newcommand{\ee}{\end{equation}}
\newcommand{\bea}{\begin{eqnarray}}
\newcommand{\eea}{\end{eqnarray}}

\title{\vspace{-1.8in}Black Hole Entropy \\ Sourced by String Winding Condensate}
\author{Ram Brustein, Yoav Zigdon}
\affil{{\normalsize Department of Physics, Ben-Gurion University, Beer-Sheva 84105, Israel} \\
{\small\tt ramyb@bgu.ac.il\ \ \ \ yoavzig@post.bgu.ac.il}}
\date{}

\begin{document}
\maketitle

\begin{abstract}
We calculate the entropy of an asymptotically Schwarzschild black hole (BH), using an effective field theory of winding modes in type II string theory. In Euclidean signature, the geometry of the black hole contains a thermal cycle which shrinks towards the horizon. The light excitations thus include, in addition to the metric and the dilaton, also the winding modes around this cycle. The winding modes condense in the near-horizon region and source the geometry of the thermal cycle. Using the effective field theory action and standard thermodynamic relations, we show that the entropy, which is also sourced by the winding modes condensate, is exactly equal to the Bekenstein-Hawking entropy of the black hole. We then discuss some properties of the winding mode condensate and end with an application of our method to an asymptotically linear-dilaton black hole.

\end{abstract}

\section{Introduction}

The idea that in string theory it is possible to calculate from first principles the Bekenstein-Hawking entropy of black holes (BHs) received the first piece of evidence in \cite{Strominger}, where the BH entropy was calculated for extremal BHs  by counting BPS states of Ramond-Ramond-charged D-branes wrapping internal cycles.

In a parallel line of thought, the idea that strings near the Hagedorn temperature \cite{AW} (see \cite{Eliezer,Mertens}, for reviews) are related to BHs and that their entropy is the BH entropy has a long history.  It was suggested by Susskind \cite{Susskind} (see also \cite{Susskind2}) that the entropy of fundamental strings with ends frozen on the BH horizon is equal to the Bekenstein-Hawking entropy of the BH.  Horowitz and Polchinski \cite{HP0},\cite{HP}, followed by Damour and Veneziano \cite{DV}, argued that a highly excited string near the Hagedorn temperature corresponds to a string scale BH, observing that the size, temperature and entropy of the string and of the BH scale the same way. See also \cite{Kutasov},\cite{Kutasov2}. In \cite{polymer} it was argued that when certain string interactions are taken into account, the correspondence could be extended to arbitrarily large BHs.

In the current paper we show that the Bekenstein-Hawking entropy of asymptotical Schwarzschild BHs of arbitrarily large size can be calculated by an effective description of closed fundamental strings wrapping a shrinking thermal cycle. Near the horizon the winding modes become light, condense and source the entropy. A partial list of papers about the phenomenon of winding modes localization near a vanishing cycle includes \cite{Atish}, \cite{Eva1},\cite{Headrick},\cite{Eva2},\linebreak \cite{Horowitz},\cite{Sunny1} \cite{Sunny2},\cite{Sunny3},\cite{Sunny4},\cite{Sunny5},\cite{Sunny6}, \cite{Mertens2}.

Dabholkar \cite{Atish} considered a winding condensate confined to the tip of an orbifold and obtained the Bekenstein-Hawking entropy by adding a Gibbons-Hawking-York term \cite{GH}.  Giveon and Itzhaki \cite{Sunny1} obtained an area law for the entropy of winding modes in the spacetime manifold $R^2 \times T^2$. Zakharov et al. \cite{Mertens2}, considered long strings near a BH horizon, and argued that the first law of thermodynamics leads to the Bekenstein-Hawking entropy for these strings if the Hagedorn temperature is redshifted to the Hawking temperature. More recently, Jafferis and Schneider \cite{Jafferis} proposed that winding strings on disconnected spacetimes in a specific entangled state, are dual to the two-sided BH (see also \cite{Sunny7}).
Chen and Maldacena discussed the winding-mode condensate in large-D BHs with horizon size $\sim D\sqrt{\alpha'}$, and mentioned the possibility that the entropy of the condensate equals that of the BH entropy \cite{Maldacena},\cite{MaldacenaStrings}. Type II stringy $\alpha'$ corrections to a BH solution and  the associated thermodynamical quantities were computed in \cite{Chen}.

The standard procedure of calculating the entropy in the classical approximation consists of first computing the on-shell action, which is given by a boundary term. Then acting with $(\beta \partial_{\beta}-1)$ on this action to obtain the entropy, $\beta$ being the inverse temperature. In this paper, we calculate the entropy by first acting with $(\beta \partial_{\beta}-1)$ on the action and then setting the fields on shell. In this case, it is not necessary to add the Gibbons-Hawking-York (GHY) term and the horizon, where the thermal cycle shrinks, is regular, free of a conical singularity \cite{Solodukhin}.

We show that this procedure leads to a formula for the entropy that can be expressed as an area integral and that the value of the entropy is exactly equal to the Bekenstein-Hawking entropy of the asymptotically flat Schwarzschild BH. We then discuss the differences between our method and the Gibbons-Hawking calculation \cite{GH} and to the introduction of a conical singularity at the horizon. We then proceed to calculate approximately the spatial distribution of the winding-mode condensate, which, as we show, is indeed confined to an invariant distance of the order of the string scale in the vicinity of the horizon. Finally, we apply our method to the ``cigar'' background \cite{cigarRef1},\cite{EdCigar},\cite{DVV} -- an asymptotically linear-dilaton stringy background.

Using instead the standard procedure for calculating the entropy,  we find that the action depends linearly on $\beta$ and consequently the entropy vanishes (see below).  Thus, if one wishes to first evaluate the action and then calculate the entropy, one needs to include another boundary term. A conventional example of the latter is the Gibbons-Hawking-York (GHY) term at infinity \cite{GH}. This is discussed in section 3.1. Another example consists of introducing a boundary term at the tip, a conical singularity, which is discussed in section 3.2.

\section{Effective field theory for winding modes}

\subsection{Action}

We consider type II closed string theory in $10$ dimensions and a background containing $D=10$-dimensional asymptotically flat Euclidean BH. The manifold contains a contractible thermal cycle that vanishes at the horizon. In the rest of the paper we keep $D$ general (but $D>3$\footnote{Although we have indications that the subsequent calculation of the entropy is valid for $2$ and $3$ dimensions as well, these cases need additional care.}), expecting that by an appropriate compactification of the critical string theory, the calculation should be valid also for $3<D < 10$ and perhaps also for $D>10$ \cite{Maldacena}.

Near the horizon, the spectrum of string theory contains light modes which wrap around the thermal cycle once. We describe these winding modes, $\chi,\chi^*$, in terms of an effective field theory. The other light fields in this theory are the dilaton $\Phi_D$, the spatial metric components $G_{\mu \nu}$ and the thermal circle component of the metric $G_{\tau \tau }\equiv e^{2\sigma}$.
We will also use the notation $\Phi_d$ for the shifted dilaton,
$
\Phi_d = \Phi_D - \frac{\sigma}{2}.
$

The action for the winding modes, including the interaction with $e^{2\sigma}$, is given by \cite{HP},
\begin{equation}\label{I1}
I_1 = \beta \int_M d^d x \sqrt{G_D} e^{-2\Phi_D} \left(G_d^{\mu \nu}\partial_{\mu} \chi\partial_{\nu} \chi^*+\frac{\beta^2 e^{2\sigma}-\beta_H ^2}{(2\pi \alpha')^2}\chi \chi^*\right),
\end{equation}
while the standard NS-NS action is given by \cite{Polchinski}:
\begin{equation}\label{I_2}
I_{2} = -\frac{\beta}{2\kappa_0^2 }\int_M d^d x \sqrt{G_D} e^{-2\Phi_D}  \left(R_d + 4G_d ^{\mu \nu}\partial_{\mu} \Phi_d \partial_{\nu}\Phi_d- G_d^{\mu \nu}\partial_{\mu} \sigma \partial_{\nu}\sigma\right).
\end{equation}
Newton's constant in the Einstein frame is related to $\kappa_0$ and the constant dilaton $\Phi_0$ in the string frame,
\begin{equation}\label{frame}
\kappa^2 \equiv  8\pi G_N = \kappa_0 ^2 e^{2\Phi_0}.
\end{equation}
The term proportional to $e^{2\sigma}$ in $I_1$,  reproduces a string S-matrix element \cite{HagedornEFT} to leading order in the string coupling.

\subsection{Equations of motion and classical action}

In this subsection we list the equations of motion (EOMs) derived from the action $I_1+I_2$. Then, we express the winding modes action as well as the total action as boundary terms.

\subsubsection{Equations of motion}

The EOMs for  $\sigma$ and $\chi$ is derived by varying the action $I_1$ with respect to $\sigma$,
\begin{equation}
\label{sigma}
\frac{e^{2\Phi_D}}{\sqrt{G_D}}\partial_{\mu} \left(\sqrt{G_D}e^{-2\Phi_D}G_d^{\mu \nu}\partial_{\nu}\sigma\right)=   \frac{\beta^2\kappa_0^2}{2\pi^2 (\alpha')^2}\chi \chi^*e^{2\sigma},
\end{equation}
and with respect to $\chi^*$,
\begin{eqnarray}\label{chi}
&\frac{e^{2\Phi_D}}{\sqrt{G_D}}\partial_{\mu} \left(\sqrt{G_D}e^{-2\Phi_D}G_d^{\mu \nu}\partial_{\nu}\chi\right)=\frac{\beta^2 e^{2\sigma}-\beta_H^2}{4\pi^2 (\alpha')^2}\chi.
\end{eqnarray}
A similar equation is obtained for $\chi^*$.
The equation of motion for the dilaton $\Phi_d$ and the spatial metric are the following,
\begin{equation}
R_D + 4\nabla^2 \Phi_D -4G^{\mu \nu}\partial_{\mu} \Phi_D \partial_{\nu} \Phi_D =2\kappa_0^2 \left[G^{\mu \nu} \partial_{\mu} \chi \partial_{\nu} \chi^* +\frac{\beta^2 e^{2\sigma}-\beta_H ^2}{(2\pi \alpha')^2}\chi \chi^*\right],
\end{equation}
and
\begin{eqnarray}\label{rr}
&R_{\mu \nu}+2\nabla _{\mu} \nabla_{\nu} \Phi_D=
&2\kappa^2  \partial_{\mu} \chi \partial_{\nu} \chi ^*.
\end{eqnarray}
We show that the action of the winding modes vanishes on-shell. Integrating by parts the winding modes action Eq. (\ref{I1}) and using the winding-mode EOM (\ref{chi}), one obtains
\begin{equation}
I_{1} = \beta \int_{\partial M} d^d x~  \sqrt{G_D} e^{-2\Phi_D} \chi^* n^{\mu}\partial_{\mu} \chi=0
\end{equation}
for any solution respecting Dirichlet boundary condition (BC) at infinity and Dirichlet/Neumann BC at the horizon. This agrees with the observation in \cite{Sunny1}.

\subsubsection{Classical action}

Next, we express the on-shell action as a boundary term. Taking into account the dilaton EOM,
\begin{equation}
I = \frac{\beta}{2\kappa_0 ^2} \int_M d^d x \sqrt{G_D}e^{-2\Phi_D}\left(4\nabla^2 \Phi_D-8 G^{\mu \nu} \partial_{\mu} \Phi_D \partial_{\nu} \Phi_D\right).
\end{equation}
Therefore the total on-shell action is given by,
\begin{equation}
\label{TotalAction}
I=-\frac{2\beta}{\kappa_0 ^2} \int_{\partial M} d^{d-1} x ~ \sqrt{G_D} n^{\mu} \partial_{\mu} \left(e^{-2\Phi_D}\right),
\end{equation}
The action therefore vanishes for a constant dilaton background and diverges for a linear dilaton background.

\section{Entropy calculation}

In this section the entropy of a black hole is computed by first taking the $\beta$ derivative of the action and then putting the fields on-shell.
We assume the following ansatz for the line element:
\begin{equation}
ds^2 = e^{2\sigma(r)} d\tau^2 +\frac{1}{g(r)} dr^2 +  r^2 d\Omega_{D-2}^2 ~,~ \tau \sim \tau + \frac{1}{T}.
\end{equation}
The metric is taken to be asymptotically flat and the dilaton tends to a constant value guaranteeing weak coupling. For a Euclidean asymptotically Schwarzschild BH, and for $r\gg r_0$,
\begin{equation}\label{ESchBH}
e^{2\sigma}=1-\left(\frac{r_0}{r}\right)^{D-3}
\end{equation}
and $g(r)=e^{2\sigma}$. Such a configuration solves the EOM asymptotically. \\
The Schwarzschild radius $r_0$ is related to the inverse temperature,
\begin{equation}\label{betar0}
\frac{\beta}{4\pi} = \frac{r_0}{D-3}.
\end{equation}
Next, the entropy can be calculated by identifying the Euclidean action $I_1+I_2$ in Eqs.~(\ref{I1}), (\ref{I_2}) as the product of the inverse temperature and the Helmholtz free energy $I_1+I_2=\beta F$. Using the standard thermodynamic relation between the free energy and the entropy,
\begin{equation}
\label{entropy}
S = \left(\beta\partial_{\beta}-1\right)I= \beta\int_M d^d x \sqrt{G_D} e^{-2\Phi_D} \frac{2\beta^2 e^{2\sigma}}{(2\pi \alpha')^2} \chi \chi^* (x),
\end{equation}
we see that the entropy density is, as expected, proportional to the magnitude of the winding-mode condensate $\chi\chi^*$. While the authors of \cite{Maldacena} considered the entropy of a free winding condensate, here we are taking into account the leading order interaction in the string coupling.
Substituting Eq.~(\ref{sigma}) yields
\begin{equation}
\label{ExtrinsicCurvature}
S = \frac{\beta}{\kappa_0^2} \int_M d^d x ~\partial_{\mu} \left(\sqrt{G_D}e^{-2\Phi_D}G_d ^{\mu \nu}\partial_{\nu}\sigma\right).
\end{equation}
Using Gauss' theorem, we find
\begin{equation}
S=\frac{\beta}{\kappa_0^2} \int_{\partial M} d^{d-1} x~ n_{\mu} \left(\sqrt{G_D}e^{-2\Phi_D}G_d^{\mu \nu}\partial_{\nu}\sigma\right).
\end{equation}

Consequently, $S$ can be expressed as a boundary integral over a sphere of radius $R_c\gg r_0$,
\begin{equation}\label{BoundaryTerm}
S=\frac{\beta}{\kappa_0^2} \int\limits_{\partial M} d^{d-1} \Omega~n_r~ \sqrt{G_{d-1}}e^{-2\Phi_D(r)}\frac{\partial_r(e^{2\sigma})}{2\sqrt{\frac{e^{2\sigma}}{g}}}.
\end{equation}
Taking the limit  $R_c\to \infty$ for an asymptotically flat D-dimensional Euclidean Schwarzschild  BH whose metric is given in Eq.~(\ref{ESchBH}),
the expression for the entropy becomes the following,
\begin{equation}
S=\frac{\beta}{\kappa_0^2} ~\int\limits_{\partial M} d^{d-1} \Omega~  e^{-2\Phi_D(\infty)}\frac{(D-3)}{2}r_0 ^{D-3}.
\end{equation}
Using Eq.~(\ref{betar0}) and Eq.~(\ref{frame}), we obtain the final answer,
\begin{equation}
S = \frac{\omega_{D-2} r_0 ^{D-2}}{4G_N},
\label{Sfinal}
\end{equation}
with $\omega_{D-2}$ being the volume of a $(D-2)$-unit sphere $\omega_{D-2}=\int d^{D-2} \Omega $.

This final value of $S$ in Eq.~(\ref{Sfinal}) is exactly equal to the Bekenstein-Hawking entropy of the BH. This is the main result of our paper.

\subsection{Comparison with the Gibbons-Hawking calculation of the black hole entropy}

Let us compare our calculation with the celebrated result of Gibbons and Hawking (GH) \cite{GH}, who calculated the Euclidean action for a Schwarzschild BH in an asymptotically flat space and pure Einstein gravity. The entropy is calculated by adding the Gibbons-Hawking-York (GHY) boundary term,  which makes the variational problem in Einstein's gravity well-posed when one wishes to fix the metric variation at the boundary. For a nice review on the variational problem in gravity, see \cite{var1}.

The GHY term is equal to the normal derivative of the area of a large constant radius $r=R_c$ surface:
\begin{eqnarray}
\label{GHY}
I_{GHY}& =&\frac{1}{8\pi G_N}\frac{\partial}{\partial n} \int d\Sigma \nonumber \\
& =&\frac{1}{8\pi G_N} \sqrt{1-\left(\frac{r_0}{r}\right)^{D-3}} \partial_r \left( \omega_{D-2} r^{D-2} \beta \sqrt{1-\left(\frac{r_0}{r}\right)^{D-3}}\right)_{\Bigg|{R_c}} \nonumber \\
& \propto& R_c^{D-3}/8\pi G_N.
\end{eqnarray}
This result is divergent and has to be regularized by adding a counter term.  Apart from requiring that the regularized expression is finite, we are not familiar with a rule that can be derived from first principles that dictate which particular counter term needs to be added. Gibbons and Hawking  chose a counter term that would make flat space at finite temperature have zero action. Therefore, they subtracted the normal derivative of the area of a constant radius flat-space surface ($d\Sigma = r^{D-2} d^{D-2}\Omega d\tau $), for which the normal and the temperature are identical to the Schwarzschild normal and temperature.

We point out a situation in which the Einstein-Hilbert action without the GHY term is well-posed.

The variation of the Einstein-Hilbert action $\delta S_{EH}$ includes varying the Ricci tensor $\delta R_{\alpha \beta}$. The Palatini identity,
$
G^{\alpha \beta}\delta R_{\alpha \beta} = \nabla_{\mu} \left[G^{\alpha \beta} \delta \Gamma^{\mu} _{\alpha \beta}-G^{\alpha \mu}\delta \Gamma^{\beta} _{\alpha \beta}\right],
$
gives rise to a boundary term in $\delta S_{EH}$. Now, instead of the conventional Dirichlet boundary condition for metric variations, one could impose the condition that this boundary term vanishes:
\begin{equation}
\label{BC}
n_{\mu} \left(G^{\alpha \beta} \delta \Gamma^{\mu} _{\alpha \beta}-G^{\alpha \mu}\delta \Gamma^{\beta} _{\alpha \beta}\right)|_{\partial M}=0.
\end{equation}
This is a single boundary condition for metrics satisfying the second-order Einstein's equations and so should allow for a family of well-behaved solutions. It combines the variation of the metric and its derivatives. In asymptotically Schwarzschild spacetime, this boundary condition is approximately a Neumann boundary condition, because $\partial_r G^{rr}\delta G_{rr} \ll G^{rr}\partial_r  \delta G_{rr}$. In  \cite{GeneralizedGravEntropy}, the same boundary condition as in Eq.~(\ref{BC}) was chosen for the metric variations.

For example, in 4D, the following (asymptotic) variations about the Schwarzschild solution, $\delta G_{\tau \tau} = \frac{A}{r^4}$ while $\delta G_{\mu \nu}=0$ for $\mu$, $\nu \ne \tau$, satisfy Eq.~(\ref{BC}) for any constant $A$ such that $\frac{A}{R_c ^4} \ll \frac{r_0}{R_c}$. More generally, there exists a linear combination of metric variations and their first derivatives about Schwarzschild, obeying Eq.~(\ref{BC}). These variations do not modify the size of the Euclidean circle at infinity, a property that was emphasized in \cite{Chen}.

Similarly to GH, we assumed that the horizon does not contribute to the entropy. This will be justified in section 4.3 by calculating explicitly the entropy of the winding-mode condensate.

\subsection{Conical singularity}

It is possible to consider the Euclidean gravitational path integral over a family of asymptotic periodicities $\beta$ \cite{Solodukhin}. For any $\beta$ such that $\beta \neq \beta_h$, where $\beta_h$ is the Hawking inverse temperature, one finds  a conical singularity at the tip. The near-horizon geometry, suppressing angular directions, is given by
\begin{equation}
ds^2 = \frac{\rho^2}{\beta_h ^2} d\tau ^2 + d\rho^2~,~ \tau \sim \tau + \beta.
\end{equation}
As proposed in \cite{Solodukhin}, in taking $\beta$-derivatives of the action, one has to consider this family and at the end of the calculation set the periodicity to the Hawking periodicity. The Ricci scalar is given by:
\begin{equation}
R = \frac{2(\beta_h-\beta)}{\beta} \delta (\rho).
\end{equation}
This produces a boundary term for the Einstein-Hilbert action
\begin{equation}
I = -\frac{2\pi}{\kappa_0^2 e^{2\Phi_H}}\text{A}_H \left(1-\frac{\beta}{\beta_h}\right).
\end{equation}
The horizon has area $A_H$ and the value of dilaton there is $\Phi_H$. Therefore,
\begin{equation}
S = \left(\beta \partial_{\beta} -1\right)I |_{\beta=\beta_h}= \frac{\text{A}_H}{4G_N}.
\end{equation}
We would like to point out that introducing a conical singularity at the tip is optional. Instead of including geometries with a conical singularity as discussed above, one can choose the Euclidean gravitational path integral to be exclusively over smooth geometries. When calculating the $\beta$-derivative of the action, one should then use two neighboring configurations, each of which with its own Hawking inverse temperature. 

\section{Some properties of the winding-mode condensate profile}

It would be interesting to find the profile for the winding-mode condensate. While we stop short of actually finding an exact expression for the profile, we can discuss some general approximate properties of the condensate profile which can be deduced from the EFT. We also perform a calculation of the entropy associated with a winding condensate in an asymptotically linear-dilaton background and find agreement with the Bekenstein-Hawking entropy.

As discussed in the introduction, we expect that the winding-modes condensate is localized in the near-horizon region near the ``tip'', where the thermal cycle shrinks and the winding modes become light.

Here, due to the approximate nature of the discussion, we restrict our attention to backgrounds for which $G^{rr}=e^{2 \sigma}$ and $\Phi_D =\text{const}$. In this case, the tip region has three significant radii. These radii as well as the value of $e^{2\sigma}$ at these radii are listed in the following equation,
\begin{equation}
\begin{aligned}
 r_0, \ \ \ & e^{2\sigma(r_0)}=0, \hspace{1.2in}  \text{Euclidean horizon}, \\
  r_\chi, \ \ \ &  e^{2\sigma(r_\chi)}=\frac{\beta_H ^2}{\beta^2}\ll 1, \hspace{.6in}  \text{ vanishing winding-mode mass}, \\
 r_V, \ \ \ & e^{2\sigma(r_V)}= 1-\frac{ r_0^{D-3}}{r_V^{D-3}}, \hspace{.2in}   \text{ vanishing winding-mode condensate} .
\end{aligned}
\label{3radii}
\end{equation}

In the region $r_0< r < r_V$ the mass-squared of the winding modes is negative according to Eq.~(\ref{I1}). Based on the analysis in \cite{HagedornEFT}, we expect that when interactions are taken into account, the mass-squared of the winding modes in this region is actually small and positive.

We further assume that in the tip region,
\begin{equation}
\partial_r e^{2\sigma} = \frac{4\pi}{\beta},
\label{dsigma}
\end{equation}
which is the value of $\partial_r e^{2\sigma}$ in the region where the condensate vanishes. This assumption will be shown to be self-consistent later, by showing that the EOMs admit such solutions.
It essentially means that in the tip region $e^{2\sigma}$ is approximately constant, meaning that  $\alpha'$ corrections in the gravity sector are negligible. It also allows us to estimate the width of the tip region,
\begin{equation}
\frac{4\pi}{\beta}(r_V-r_0)= \frac{\beta_H^2}{\beta^2},
\end{equation}
So,
\begin{equation}
r_V-r_0 = \frac{\beta_H^2}{4\pi\beta},
\end{equation}
which corresponds to an invariant distance of the order of the string length,
\begin{equation}
\sqrt{g_{rr}(r_V-r_0)^2} =\frac{\beta}{\beta_H} \frac{\beta_H^2}{4\pi\beta} =\sqrt{\frac{\alpha'}{2}}.
\end{equation}

As a first step, we can  parameterize the profile as follows:
\begin{equation}
\label{Profile}
\chi \chi^* (r) = A \delta(r-r_\chi).
\end{equation}
The Dirac $\delta$-function should be thought of as a limiting version of some regular function which is highly peaked near the tip with a width of the order of the string length. Substituting this form into Eq.~(\ref{entropy}) and approximating $e^{2\sigma}\approx \beta_H ^2/\beta^2$,  we find that $A=\dfrac{\pi \alpha'}{2\beta \kappa_0^2 }\left(\frac{r_0}{r_{\chi}}\right)^{D-2}\simeq \dfrac{\pi \alpha'}{2\beta \kappa_0^2 }$, reproduces the BH entropy.

In \cite{Sunny2}, for a different background (See Sect.~(4.3)), the profile was found to be a Gaussian in the radial Rindler coordinate, which corresponds to an exponential in Schwarzschild coordinates . This function belongs to the set of regularized delta-functions, so its limit as $\alpha'\to 0$ is consistent with Eq.~(\ref{Profile}). In \cite{Mertens2}, similarly, the profile was found to be a Gaussian multiplied by a Laguerre polynomial.

To resolve the profile in more detail we need to solve the winding mode equation of motion, which we do next.

\subsection{Approximate solution of the winding-mode equation of motion }

Eq.~(\ref{chi}) and the assumptions of spherical symmetry, constant dilaton, $g(r)=e^{2\sigma}$, is given by
 \begin{equation}
 \frac{1}{r^{D-2}}\partial_r\left(r^{D-2}e^{2\sigma}\partial_r\chi(r)\right)=\frac{\beta^2 e^{2\sigma(r)}-\beta_H ^2}{(2\pi \alpha')^2}\chi(r).
 \end{equation}
In the vicinity of $r_{\chi}$, $e^{2\sigma(r)}\approx a \frac{\beta_H ^2}{\beta^2}$ where $a$ is a dimensionless number of order unity. Using Eqs.~(\ref{3radii}), (\ref{dsigma}), the EOM of $\chi$ can be approximated as,
\begin{equation}
\label{ChiAprroxEOM}
 a\frac{\beta_H ^2}{\beta^2} \partial_r ^2 \chi+\left(\frac{D-2}{r_{\chi}}+\frac{4\pi}{\beta}\right)\partial_r \chi = \frac{\beta}{\pi \alpha'^2} (r-r_{\chi})\chi.
\end{equation}
Since $r_{\chi}- r_0 \ll r_0$, it follows that $r_\chi\approx  \frac{(D-3) \beta}{ 4 \pi}$. Then, the equation simplifies,
\begin{equation}
a\partial_r ^2 \chi + \frac{2D-5}{D-3} \frac{\beta}{2\pi \alpha'}\partial_r \chi = \left(\frac{\beta}{2\pi \alpha'} \right)^3(r-r_{\chi}) \chi.
\label{ChiAprroxEOM1}
\end{equation}
The solution is approximately constant for $r-r_\chi \lesssim \frac{2\pi \alpha'}{\beta}$, which, as we recall, corresponds to string length invariant distance and decreases exponentially for $r-r_\chi \gg \frac{2\pi \alpha'}{\beta}$. This indicates that $r_V-r_{\chi}\simeq \frac{2\pi \alpha'}{\beta}$.

The decreasing solution of Eq.~(\ref{ChiAprroxEOM1}) can be expressed in terms of the Airy function
\begin{equation}\label{ProfileResolved}
\chi(r) \propto  e^{-\frac{2D-5}{2(D-3)}\frac{\beta (r-r_{\chi})}{2\pi a \alpha'}}Ai\left(\frac{\beta (r-r_{\chi})}{2\pi a \alpha'}+\left(\frac{2D-5}{2a(D-3)}\right)^2\right).
\end{equation}
The Airy function decays slower than the exponential factor. It follows that $\chi$ decreases by a factor $1/e$ from its value at $r_\chi$ for
\begin{equation}
r-r_{\chi} =\frac{2(D-3)}{2D-5}\frac{2\pi a \alpha'}{\beta}.
\end{equation}
This amounts to an invariant distance $\frac{2(D-3)}{2D-5}\sqrt{\frac{a\alpha'}{2}}$, which is of order of the string scale.

In summary, we found that the condensate decays, roughly, over an invariant distance of order of a few string lengths away from the horizon. The relation to the delta-function in Eq. (\ref{Profile}) is established if one normalizes the profile and considers the width of the profile to be parametrically small $\frac{\sqrt{\alpha'}}{r_0}\to 0$.

\subsection{Weakly-curved near-horizon region }

We wish to point out that even though we discussed string length invariant distances, the EFT remains valid in the near-horizon region $r>r_{\chi}$. The reason is that, as explained below, curvatures and the derivatives of the condensate are small is string units.

The Ricci scalar is given by
\begin{equation}
R_D=-\frac{1}{r^{D-2}}\partial_r^2 \left(r^{D-2}e^{2\sigma} \right) +  \frac{(D-2)(D-3)}{r^2}.
\end{equation}
The non-vanishing components of the Ricci tensor, $R^\tau_{\ \tau}=R^r_{\ r}$ and $R^{\alpha}_{\ \alpha}$, $\alpha \ne r, \tau$ are the following,
\begin{eqnarray}
R^r_{\ r} &=& -\frac{1}{2}\frac{1}{r^{D-2}}\partial_r \left(r^{D-2} \partial_r e^{2\sigma} \right), \\
R^{\alpha}_{\ \alpha} &=& \tfrac{D-3}{r^2}-\frac{1}{r^{D-2}}\partial_r\left(r^{D-3}  e^{2\sigma} \right).
\end{eqnarray}

Recalling Eqs.~(\ref{3radii}), (\ref{dsigma}), it follows that in the near-horizon region all curvatures are of the order of  $\frac{1}{r_0^2}$. In this case, it is justified to neglect such low curvatures in the near-horizon region.

Another quantity which has to be small for the background to be weakly-curved is $\frac{G^{rr}(r-r_{\chi})\partial_r \chi  }{\chi} $ (also higher derivatives should be small.) Indeed, Eq. (\ref{3radii}) and (\ref{ProfileResolved}) imply that it is of order $\frac{\beta_H ^2}{\beta^2} \ll 1$.

\subsection{Cigar Background}

Until now we discussed asymptotically constant dilaton BHs. Here we show that our methods
can also be applied successfully to calculate the entropy of the BH background associated with the near-horizon limit of $k$ near-extremal NS5 branes \cite{MaldacenaNS5}. This is specified by the following line element, dilaton and asymptotic inverse temperature,
\begin{equation}\label{cigar1}
ds^2 = \tanh^2 \left(\frac{\rho}{\sqrt{\alpha' k}}\right)d\tau ^2+d\rho^2 +\alpha' k d\Omega^2 _3+(dy_1 ^2 +...+dy_5 ^2),
\end{equation}
\begin{equation}\label{cigar2}
e^{-2\Phi_D} = \frac{1}{g_0 ^2} {\cosh^2 \left(\frac{\rho}{\sqrt{k \alpha'}}\right)},
\end{equation}
\begin{equation}\label{cigar3}
\beta = 2\pi \sqrt{k\alpha'}.
\end{equation}
For $k\gg 1$, the background is weakly curved.

The winding profile for this background for large $k$ was found by Giveon and Itzhaki \cite{Sunny2} by observing that the effective mass-squared of the winding modes near the tip in Rindler coordinates behaves like a harmonic oscillator potential. In this case, the condensate ground state wavefunction is a Gaussian in the radial Rindler coordinate.
The profile, normalized by $\frac{1}{\kappa_0}$, is given by
\begin{equation}
\chi(\rho)= \frac{1}{\kappa_0}{e^{-\tfrac{\rho^2}{2\alpha'}}}.
\label{cigarprof}
\end{equation}

Substituting Eqs.~ (\ref{cigar1}),(\ref{cigar2}) and (\ref{cigar3}) into Eq.~(\ref{entropy}) yields
\begin{eqnarray}
S &=& \frac{2(2\pi \sqrt{k\alpha'})^3}{(2\pi \alpha')^2 g_0 ^2\kappa_0 ^2}\int_0 ^{\infty} d\rho ~e^{-\frac{\rho^2}{\alpha'}}\tanh^3\left(\frac{\rho}{\sqrt{k \alpha'}}\right)\cosh^2 \left(\frac{\rho}{\sqrt{k \alpha'}}\right) \cr \vspace{.4in} &\times& \text{Vol}(T^5) (\alpha' k)^{\frac{3}{2}}\omega_3. \ \
\end{eqnarray}
For $k\gg 1$, $\tanh(\frac{\rho}{\sqrt{k \alpha'}})\approx \frac{\rho}{\sqrt{k \alpha'}}$. As a result, in this limit,
\begin{eqnarray}
S &=& \frac{2 (2\pi \sqrt{\alpha ' k})^3}{(2\pi \alpha')^2 g_0 ^2 \kappa_0 ^2}\int_0 ^{\infty} \frac{\rho^3}{(\sqrt{k \alpha'})^3}e^{-\frac{\rho^2}{\alpha'}}d\rho \times \text{Vol}(T^5) (\alpha' k)^{\frac{3}{2}}\omega_3 \cr
&=& \frac{2\pi }{g_0 ^2 \kappa_0 ^2}\times \text{Vol}(T^5) (\alpha' k)^{\frac{3}{2}}\omega_3=\frac{{A_H}}{4G_N},
\end{eqnarray}
with $A_H$ being the horizon area of the BH.

As another check, the background equations Eqs.~(\ref{cigar1}),(\ref{cigar2}) and (\ref{cigar3}), can be substituted into Eq.~(\ref{BoundaryTerm}), resulting in exactly the same Bekenstein-Hawking entropy.

To summarize, the cigar background with the Gaussian profile in Eq.~(\ref{cigarprof}) does lead to the correct Bekenstein-Hawking entropy of the BH.

\section{Summary and Outlook}

We demonstrated that the BH entropy can be reproduced by an effective description of winding modes living near the horizon of a Euclidean BH. Our calculation appears similar in spirit to the Gibbons-Hawking calculation because it involves a boundary term.
However, it is distinct from the Gibbons-Hawking result in that it does not require invoking a particular regularization scheme. We then proceeded to calculate approximately the profile of the winding modes in the near-horizon region and found that, as expected, the condensate decays within a few invariant string lengths.

Some interesting future improvement of our analysis would be to prove that there is a string background, similar in form to the approximate background that we found, by solving all the EOMs, taking into account the back-reaction of the non-trivial profile of the winding strings.

An even more interesting improvement would consist of finding a microscopic description of the winding strings, which may allow a precise counting of microscopic stringy degrees of freedom, for instance, by counting  the number of string bits in a long winding string.

Extending the present calculation to asymptotically AdS BHs is interesting too; in this case we expect that  one would need to subtract appropriate counter terms to get the Bekenstein-Hawking entropy. The dual conformal field theory degrees of freedom of the winding modes may help in microscopically counting the entropy.

\section*{Acknowledgement}

We would like to thank Ohad Mamroud, Eran Palti, Yotam Sherf and Erez Urbach for useful discussions.
We are especially grateful to Sunny Itzhaki for useful comments and discussions.
We thank Yiming Chen, Juan Maldacena, Joey Medved and Thomas Mertens for comments on the manuscript. 
The research was supported by the Israel Science Foundation grant no. 1294/16.

\end{document}